\documentclass[notitlepage,twocolumn,letterpaper,natbib,aps,prl,amsmath,amsfonts,nofootinbib,preprintnumbers,superscriptaddress,secnumarabic]{revtex4-1}

\usepackage{amssymb,amsmath,latexsym,mathrsfs}
\usepackage{graphicx}
\usepackage{physics}
\usepackage{mathtools}
\usepackage{bm}% bold math
\usepackage{bbold}
\usepackage[usenames,dvipsnames]{color}
\usepackage[breaklinks,colorlinks,urlcolor=magenta,citecolor=magenta,linkcolor=magenta]{hyperref}
\usepackage{xcolor}

\begin{document}

\title{Superluminal signalling witness for quantum state reduction}

\author{Aritro Mukherjee}
\affiliation{Institute for Theoretical Physics Amsterdam,
University of Amsterdam, Science Park 904, 1098 XH Amsterdam, The Netherlands}
\author{Lisa Lenstra}
\affiliation{Institute for Theoretical Physics Amsterdam,
University of Amsterdam, Science Park 904, 1098 XH Amsterdam, The Netherlands}
\author{Lotte Mertens}
\affiliation{Institute for Theoretical Physics Amsterdam,
University of Amsterdam, Science Park 904, 1098 XH Amsterdam, The Netherlands}
\author{Jasper van Wezel}
\affiliation{Institute for Theoretical Physics Amsterdam,
University of Amsterdam, Science Park 904, 1098 XH Amsterdam, The Netherlands}

%%%%%%%%%%%%%%%%%%%%%%%%%%%%%%%%

\begin{abstract}
Models for quantum state reduction address the quantum measurement problem by suggesting weak modifications to Schr\"odinger's equation that have no observable effect at microscopic scales, but dominate the dynamics of macroscopic objects. Enforcing linearity of the master equation for such models has long been used as a way of ensuring that modifications to Schr\"odinger's equation do not introduce a possibility for superluminal signalling. In large classes of quantum state reduction models, however, and in particular in models employing correlated noise, formulating a master equation for the quantum state is prohibitively difficult or impossible. Here, we formulate a witness for superluminal signalling that is applicable to generic quantum state reduction models, including those involving correlated as well as uncorrelated noise. Surprisingly, application of the witness to known models described by linear master equations shows that these may still admit superluminal signalling, unless a particular locality condition is obeyed. In contrast, we show that the witness introduced here provides a necessary and sufficient condition for excluding superluminal signals under all circumstances. We further apply the witness to several models driven by physical, correlated noise, where linear master equations are not analytically obtainable, and find that they allow for superluminal signalling. We suggest how specific correlated-noise models may be able to avoid it, and that the witness introduced here provides a stringent guide to constructing such models.
\end{abstract}

\maketitle

%%%%%%%%%%%%%%%%%%%%%%%%%%%%%%%%

~\noindent\emph{\textbf{Introduction} ---}
Quantum mechanics has been shown to be non-local by the violation of Bell inequalities in (loophole-free) Bell test experiments~\cite{Bell_thm,Bell_test_1,Bell_test_2,Bell_test_3,Bell_test_4,Bell_test_5,Bell_test_6}. Despite the non-local evolution of quantum states evidenced by the violations of Bell inequalities, the Bell tests are well-known not to allow faster-than-light communication or superluminal signalling (SLS)~\cite{Bell_FTL,teleport_1,teleport_2,teleport_3}. The projective and probabilistic nature of quantum measurements renders the local observers in Bell test setups incapable of distinguishing the effect of measurements or quantum operations performed at causally disconnected events.

Quantum measurement itself, meanwhile, remains one of the central open problems of modern physics, being at odds with the unitary time evolution prescribed by Schr\"odinger's equation~\cite{measurement_problem_1,measurement_problem_2,measurement_problem_3,measurement_problem_4}. Models for dynamical quantum state reduction (DQSR) attempt to resolve this dichotomy by proposing alterations of Schr\"odinger's equation that are weak enough to allow the time evolution of microscopic objects to be practically indistinguishable from unitary quantum dynamics, while their collective effect on macroscopic objects such as measurement devices causes their quantum state to reduce to a classical state~\cite{measurement_problem_4, Bassi_03_PhyRep,percival1998,symmetry,vanwezelprb}. In order for DQSR to be able to result in distinct and stable macroscopic measurement outcomes, starting from the measurement of identical microscopic quantum states, models of DQSR are necessarily stochastic and non-unitary. To obtain the correct Born rule statistics of measurement outcomes, they moreover need to be non-linear~\cite{mertenspra}.

Although the non-unitary and non-linear nature of DQSR models imply that in general they will allow for superluminal signalling, this is avoided in particular DQSR models. A widely used criterion for identifying such models is to insist that their ensemble averaged state evolution is described by linear quantum semi-group dynamics, in the form of a Gorini-Kosakowsky-Sudarshan-Lindblad (GKSL) type master equation~\cite{Bassi_03_PhyRep,Bassi2015,Gisin:1989sx}. The practical use of this condition, however, is hampered by the fact that explicit expressions for the ensemble averaged state evolution are often prohibitively hard to formulate. In particular, DQSR models employing correlated (colored) noise as their stochastic component rarely allow for closed-form master equations without making further approximations~\cite{Aritro_PRA,Bassi_03_PhyRep,cCSL1}. 

Here, we therefore introduce an alternative condition for avoiding SLS, in the form of a `witness': a quantity whose non-zero instantaneous expectation value indicates SLS. The witness can be straightforwardly evaluated in any modified quantum theory or DQSR model, including those based on correlated as well as uncorrelated (white) noise. While we show that the witness presents both a necessary and sufficient condition for avoiding SLS under all circumstances, we find that having a linear, GKSL type master equation is not sufficient to guarantee the absence of SLS. The witness allows us to introduce a further locality condition that can be imposed on the master equation in addition to linearity, which prevents signalling in all cases considered. We further demonstrate the use of the witness in cases where a master equation is not available, by analysing several recently-proposed DQSR models based on physical, colored noise dynamics~\cite{Aritro_Entropy,mertenscore}. We find that the witness indicates SLS is generically allowed in these models and we discuss the limits in which it may be avoided. The witness introduced here thus provides stringent constraints on the construction of DQSR models, guiding in particular the further exploration of models based on correlated noise.

%%%%%%%%%%%%%%%%%%%%%%%%%%%%%%%%

~\noindent\emph{\textbf{SLS witness} ---}
All models for DQSR have a limit in which they produce instantaneous projective measurement, with outcomes distributed according to Born's rule. This is usually the thermodynamic limit of infinitely large or heavy measurement machines. DQSR models smoothly connect the instantaneous projective dynamics of this limit to the continuous unitary evolution of Schr\"odinger's equation that is observed in the opposite, microscopic limit of small or light objects. This implies the existence of a mesoscopic regime, in which evolution is neither unitary, nor an instantaneous collapse. It is the existence of this mesoscopic regime that potentially allows for SLS.

To create superluminal signals, one requires a sender (say Alice) who is space-like separated from a recipient (Bob), with whom she shares a known entangled state. Entanglement is a necessary resource, because relativistic classical physics does not allow SLS, while product states contain only local information for each observer. Alice and Bob can potentially send signals using DQSR dynamics if at least one of them operates in the mesoscopic regime where DQSR deviates from both unitary evolution and instantaneous projections. We therefore consider the situation in which Alice performs a measurement on her half of the entangled state using a mesoscopic measurement device. This causes non-unitary DQSR dynamics lasting a finite, non-zero amount of time before resulting in a stable measurement outcome and final state. During or after this time, Bob can perform an instantaneous, projective measurement of his half of the entangled state using a truly macroscopic measurement apparatus. In the discussion section we will separately consider the situation in which both Alice and Bob employ a mesoscopic device, and show that it does not yield opportunities for SLS beyond those already present if only Alice uses a mesoscopic device. 

Without loss of generality, we consider a two-state superposition and write the initial configuration as $\ket{M_A} (\alpha(0) \ket{0}_A \ket{0}_B + \beta(0) \ket{1}_A \ket{1}_B ) \ket{M_B}$. Here $\ket{M_{A,B}}$ are states for the measurement devices of Alice and Bob, while the states labeled by $A$ ($B$) subscripts describe Alice's (Bob's) part of the shared entangled quantum state. As Alice performs a measurement with her mesoscopic device, but before Bob performs any measurement, the system evolves to~\cite{Von_Neumann2018-bo,Bassi_03_PhyRep,measurement_problem_4}:
\begin{align}
\left[ \, \alpha(\xi_t,t) \ket{M_A^0(t)} \ket{0(\xi_t,t)}_A \ket{0}_B \right. \phantom{]} &\notag\\
\left. + \, \beta(\xi_t,t) \ket{M_A^1(t)} \ket{1(\xi_t,t)}_A \ket{1}_B \, \right] & \ket{M_B}
\label{state}
\end{align}
Here $| M_A^{0,1}(t) \rangle$ indicates the state of the measurement device evolving differently in the two components of the wave function. Notice that the non-linear DQSR dynamics describing Alice's measurement is assumed to act locally on the entangled state, influencing only Alice's half and not Bob's. This means that diagonal elements in the time evolution operator (proportional to $\hat{\sigma}^z_A\otimes\hat{I}^{\phantom{z}}_B$) cause evolution of the coefficients $\alpha$ and $\beta$, while off-diagonal elements (proportional to $\hat{\sigma}^{x,y}_A\otimes\hat{I}^{\phantom{z}}_B$) cause transitions of Alice's state between $\ket{0}_A$ and $\ket{1}_A$. The latter effects cannot be described by evolution of the coefficients and are included in Eq.~\eqref{state} by using a time-dependent definition of Alice's basis states. The evolutions of both the coefficients and the basis states depend not only on time, but also on the (time-dependent) values of stochastic parameter(s) $\xi_t$ appearing in the particular DQSR model under consideration. That is, DQSR models are defined in terms of stochastic differential equations, in which the evolution of the quantum state at time $t$ depends on the instantaneous value(s) of a (set of) stochastic parameters(s) $\xi_t$~\cite{measurement_problem_4, Bassi_03_PhyRep,percival1998,Aritro_PRA}. Notice however, that the stochastic parameters $\xi_t$ themselves evolve in time independently from the quantum state. In each realisation of the DQSR dynamics, the outcome is determined by the particular instance of $\xi_t$ encountered, and stochastic expectation values, denoted by $\mathbb{E}_\xi[\,\cdot\,]$, are defined as the average over an ensemble of realisations of the stochastic parameter(s). For simplicity, we ignore the Schr\"odinger evolution of Bob's quantum state (i.e. we work in a co-rotating frame).

At some point during the evolution described by Eq.~\eqref{state}, Bob performs a projective measurement and instantaneously reduces the state of the full system to either $\ket{M_A^0(t)} \ket{0(\xi_t,t)}_A \ket{0}_B \ket{M_B^0}$, with probability $|\alpha(\xi_t,t)|^2$, or to $\ket{M_A^1(t)} \ket{1(\xi_t,t)}_A \ket{1}_B \ket{M_B^1}$, with probability $|\beta(\xi_t,t)|^2$. Because Bob's projective dynamics is probabilistic, he will not be able to glean any information from the one measurement outcome obtained in just a single measurement. We will thus allow for Alice and Bob to share an infinitely large ensemble of known, identically prepared entangled states. Moreover, we assume that both Alice and Bob can perform simultaneous but independent measurements on all their states, so that they can measure ensemble averages and determine outcome probabilities within the time required for a single measurement. 

If Bob can decide, based on his own measurement outcomes, whether or not Alice started performing measurements on her side, this constitutes signalling across space-like separated events. Formally, SLS is therefore avoided if and only if:
\begin{align}
\mathbb{E}_\xi\left[\left. \langle\hat{I}^A\otimes\hat{O}^B\rangle \right| \mathcal{M}^A_\xi \right] = \mathbb{E}_\xi\left[\left. \langle\hat{I}^A\otimes\hat{O}^B\rangle \right| \mathcal{I}^A \right] \,~ \forall \, \hat{O}^B
\label{SLS_condition}
\end{align}
Here, $\mathbb{E}_\xi[ \langle \hat{O} \rangle | \mathcal{M}]$ indicates the ensemble average of measurement outcomes when Bob measures observable $\hat{O}$, conditional on Alice evolving her state according to the map $\mathcal{M}$. Notice that this involves both an average over all possible paths the stochastic parameters ($\xi_t$) in Alice's DQSR evolution, and the usual quantum expectation value averaging over possible outcomes of Bob's projective measurement. The operator $\hat{I}^A\otimes\hat{O}^B$ has support only on Bob's half of the entangled state and his device, while $\mathcal{M}^A_\xi$ indicates evolution according to a DQSR model with stochastic parameter $\xi_t$ caused by Alice's mesoscopic measurement device, and the identity map $\mathcal{I}^A$ indicates Alice does not perform any measurement or operation. 

If the two expectation values in Eq.~\eqref{SLS_condition} are equal for all possible local observables $\hat{I}^A\otimes\hat{O}^B$ and all possible initial configurations, it is impossible for Bob to distinguish between Alice performing a measurement using DQSR dynamics and Alice not doing anything at all. In contrast, the violation of Eq.~\eqref{SLS_condition} for even a single observable or initial state implies the possibility of superluminal signalling between Alice and Bob. 

Writing the initial entangled quantum state as the Schmidt decomposition $\sum_j \alpha_j \ket{j}_A \ket{j}_B$, substituting this form into Eq.~\eqref{SLS_condition}, and using the fact that the equality has to hold for all observables $\hat{O}_B$, the condition of Eq.~\eqref{SLS_condition} takes on the form of a particularly convenient witness:
\begin{align}
\mathbb{E}_\xi\left[ \left| \alpha_j(\xi_t,t) \right|^2 \right] - \left| \alpha_j(0) \right|^2 = 0 ~~ \forall j, t
\label{martingale}
\end{align}
As long as this equation is satisfied, Bob cannot use local measurements to decide whether or not Alice performed a measurement. Conversely, whenever the ensemble averaged value of any coefficient in the Schmidt basis changes as a consequence of Alice's DQSR evolution, Bob can measure this using projective measurements.

%%%%%%%%%%%%%%

~\noindent\emph{\textbf{No signalling and the Martingale condition} ---} To see explicitly how superluminal communication can be achieved in cases where the witness of Eq.~\eqref{martingale} is not zero, first consider the extreme limit in which both Alice and Bob employ instantaneous projective measurements. In that case the initial state $\ket{M_A} (\alpha(0) \ket{0}_A \ket{0}_B + \beta(0) \ket{1}_A \ket{1}_B ) \ket{M_B}$ will be instantaneously reduced by Alice to either $\ket{M_A^0} \ket{0}_A \ket{0}_B \ket{M_B}$, with probability $|\alpha(0)|^2$, or to $\ket{M_A^1} \ket{1}_A \ket{1}_B \ket{M_B}$, with probability $|\beta(0)|^2$. The consecutive measurement by Bob will with certainty produce either the state $\ket{M_B^0}$ or $\ket{M_B^1}$ for Bob's measurement machine. The overall probability for Bob to register the result $\ket{M_B^0}$ is thus $|\alpha(0)|^2$, which is the same as it would have been had Bob acted directly on the initial state without Alice doing anything. Notice that adding the possibility for Alice to act with local unitary operators on her state does not affect this result, as these do not change the values of the coefficients in the wave function, nor the state at Bob's side. Under arbitrary Schr\"odinger evolution and instantaneous collapse with Born rule probabilities, the witness thus remains zero and Bob cannot distinguish between Alice performing a measurement or not.

Next, consider the situation in which Alice causes evolution according to some DQSR model, but with her evolution reaching a final state before Bob does his measurement. If the DQSR model correctly describes evolution ending on stable measurement outcomes, the state just before Bob initiates his measurement will be either $\ket{M_A^0} \ket{0(\xi_t,t)}_A \ket{0}_B \ket{M_B}$, with some probability $P_0(\xi)$, or $\ket{M_A^1} \ket{1(\xi_t,t)}_A \ket{1}_B \ket{M_B}$, with probability $P_1=1-P_0$. As before, Bob's measurement outcome can be predicted by Alice with certainty in either case, and Bob registers $\mathbb{E}_\xi[P_0(\xi)]$ as the overall probability for obtaining the outcome $\ket{M_B^0}$. Because Alice and Bob are allowed to know the initial state (for example because they prepared it together in the past), Bob will know that Alice acted with DQSR dynamics whenever $\mathbb{E}_\xi[P_0(\xi)] \neq |\alpha(0)|^2$ and the witness is nonzero. To avoid SLS, any model for DQSR must thus ensure that within an ensemble of stochastic evolutions, the stable end states representing measurement outcomes are obtained with Born rule probabilities. Moreover, this must hold for all possible initial states.

Finally, consider the case of Alice causing DQSR dynamics, and Bob imposing a projective measurement before Alice's device reaches a stable measurement outcome. The state just before Bob measures is then given by Eq.~\eqref{state}, and Bob obtains the outcome $\ket{M_B^1}$ with probability $\mathbb{E}_\xi[ \,|\alpha(\xi_t,t)|^2]$. Again, Bob can compare this probability with the known value $|\alpha(0)|^2$ that he would obtain if Alice does nothing. In order to avoid SLS, the ensemble averaged squared values of components in the Schmidt basis must therefore always remain equal to their initial values: $\mathbb{E}_\xi[ |\alpha_j(\xi_t,t)|^2 ] = |\alpha_j(0)|^2$ for all components $j$ and all times $t$. This condition renders (marginalized) probability dynamics under SLS-avoiding DQSR evolution a so-called Martingale process~\cite{Aritro_PRA,Ghirardi_90_PRA,revuz1999continuous, oksendal2003stochastic}, and shows that deviations of DQSR dynamics from being a Martingale process serve as a witness for SLS.

%%%%%%%%%%%%%%%%

~\noindent\emph{\textbf{A necessary and sufficient condition} ---}
So far, we proved the condition of Eq.~\eqref{martingale} to be necessary for guaranteeing the absence of superluminal signalling, by formulating an explicit protocol causing SLS when it is violated. That the condition is also sufficient follows from the fact that Eq.~\eqref{martingale} guarantees the reduced density matrix for Bob's half of the entangled state and his measurement machine to be independent of time. This is straightforwardly confirmed by constructing the full density matrix from Eq.~\eqref{state}, taking the partial trace over Alice's degrees of freedom, and substituting Eq.~\eqref{martingale}. As long as Bob's reduced density matrix is time-independent, he cannot observe any time-evolution in the full state of Eq.~\eqref{state} using local measurements. Conversely, if any of the time evolution in the full state would cause evolution of Bob's reduced density matrix, this would be directly observable. The witness condition of Eq.~\eqref{martingale} thus provides both a necessary and sufficient condition to avoid superluminal signalling in any dynamical evolution of the form of Eq.~\eqref{state}. The only assumptions used, are that the DQSR dynamics of the full quantum state initiated by Alice acts locally and that Bob enacts instantaneous projective measurements.

This leaves only the most general case, of Bob and Alice both using DQSR dynamics, to be considered. Again, if either Alice or Bob can decide based on their own measurement outcomes whether or not the other started measuring, this would constitute superluminal signalling. Analogous to Eq.~\eqref{SLS_condition}, the condition for Bob to be ignorant of Alice's actions is given by $\mathbb{E}_{\xi_A,\xi_B}\left[\left. \langle\hat{I}^A\otimes\hat{O}^B\rangle \right| \mathcal{M}^A_{\xi_A},\mathcal{M}^B_{\xi_B} \right]=
\mathbb{E}_{\xi_B}\left[\left. \langle\hat{I}^A\otimes\hat{O}^B\rangle \right| \mathcal{I}^A,\mathcal{M}^B_{\xi_B} \right]$ for all $\hat{O}^B$. Here, we labelled the stochastic parameters driving the DQSR mechanisms of Alice and Bob by $\xi_A$ and $\xi_B$ respectively. On her side, Alice remains unaware of whether Bob measures anything if $\mathbb{E}_{\xi_A,\xi_B}\left[\left. \langle\hat{O}^A\otimes\hat{I}^B\rangle \right| \mathcal{M}^A_{\xi_A},\mathcal{M}^B_{\xi_B} \right]=\mathbb{E}_{\xi_A}\left[\left. \langle\hat{O}^A\otimes\hat{I}^B\rangle \right|  \mathcal{M}^A_{\xi_A},\mathcal{I}^B\right]$ for all $\hat{O}^A$. Moreover, for the measurement dynamics to be free of SLS, it should also still obey Eq.~\eqref{SLS_condition}, and a similar equality in terms of Alice's observable. Expressing all these conditions in terms of the Schmidt decomposition $\sum_j \alpha_j \ket{j}_A \ket{j}_B$ for the entangled state, we find that SLS is avoided in the fully general case if and only if:
\begin{align}
\mathbb{E}_{\xi_A,\xi_B}\left[\,|\alpha_j (\xi_A, \xi_B, t)|^2\right] = \mathbb{E}_{\xi_B}\left[\,|\alpha_j (\xi_B, t)|^2\right] \notag \\
= \mathbb{E}_{\xi_A}\left[\,|\alpha_j (\xi_A, t)|^2\right] = \left| \alpha_j(0) \right|^2 ~~~ \forall j,t \label{generalMartingale}
\end{align}
This hierarchy of conditions is a direct generalization of Eq.~\eqref{martingale}, where instead of a single stochastic parameter, it is now required that marginalizing over any possible set of stochastic parameters controlling the DQSR dynamics of any parties involved, always yields a Martingale process. Deviations of DQSR dynamics from being a Martingale process in this general sense serve as witnesses for SLS in the most general setting.

The condition of Eq.~\eqref{generalMartingale} simplifies if the DQSR dynamics employed by Alice and Bob acts only on their own device and their own part of the shared entangled state. In that case, the evolution within an infinitesimal time step can be thought of as the DQSR dynamics caused by Alice and Bob occurring separately and successively. As a result, two conditions of the type of Eq.~\eqref{martingale}, one in terms of $\xi_A$ and one with $\xi_B$, suffice to ensure that the ensemble average $\mathbb{E}_{\xi_A,\xi_B}\left[\,|\alpha_j (\xi_A, \xi_B, t)|^2\right]$ as well as the expectation values of all local observables remain constant during the infinitesimal time step. Because the same argument can be repeated for any subsequent time step, adherence to Eq.~\eqref{martingale} is a necessary and sufficient condition to preclude superluminal signalling in this case.

%%%%%%%%%%%%%%%%%%%%%

~\noindent\emph{\textbf{Linearity is not a sufficient condition} ---}
It has been previously shown that linearity of the master equation, if it is available, is a necessary condition for avoiding SLS~\cite{Gisin:1989sx,Bassi2015}. We can now use Eq.~\eqref{martingale} to show that this condition does not by itself exclude all possibilities of signalling, and is thus not a sufficient condition. 

To see this, consider again the scenario where Alice enacts DQSR dynamics while Bob initially does nothing, as in Eq.~\eqref{state}. We will assume that the DQSR dynamics admits a quantum linear semi-group of the GKSL form (acting on states within the full shared Hilbert space of Alice and Bob) for an ensemble of realisations of Alice's dynamics. In that case, we can write the noise averaged expectation value of any operator $\hat{O}$ as $\mathbb{E}_\xi[\langle\hat{O}\rangle|\mathcal{M}_\xi^A] = Tr[\hat{O}\hat{\rho}]$, where the full density matrix $\hat{\rho}$ evolves via:
\begin{align}
\label{GKSL}
    \partial_t \hat{\rho} &=-i[\hat{H},\hat{\rho}]+\sum_j \Gamma_j\left( \hat{L}_j\hat{\rho}\hat{L}_j^\dagger-\frac{1}{2}\{\hat{L}_j^\dagger\hat{L}_j,\hat{\rho}\} \right) \notag \\
    &\equiv \Lambda[ \hat{\rho}].
\end{align}
We will focus on the simplest case, with $\hat{H}=0$ and $\hat{\rho}$ being initially a pure states of the form $\hat{\rho}=\ket{\psi}\bra{\psi}$, with $\ket{\psi}=\sum_{i,j} \psi_{i,j} \ket{i}_A\ket{j}_B$ an arbitrary entangled state. We will show that this simplest case allows for SLS under the linear GKSL dynamics. The more general situation of a mixed initial density matrix will then also allow SLS, as it is a convex sums of pure states.

First note that for any local observable $\hat{O}=\hat{I}^A\otimes\hat{O}^B$ appearing in Eq.~\eqref{SLS_condition}, we have $\Tr[\hat{O}\hat{\rho}] = \Tr_B[\hat{O}^B\hat{\rho}_B]$, where $\hat{\rho}_B= \Tr_A[\hat{\rho}]$ is the reduced density matrix corresponding to Bob. Using Eq.~\eqref{SLS_condition}, SLS is avoided if $\Tr_B[\hat{O}^B \hat{\rho}_B(t)] = \Tr_B[\hat{O}^B \hat{\rho}_B(0)]~ \forall t$. Since this must hold for any choice of $\hat{O}_B$, it implies $\partial_t \hat{\rho}_B (t)= 0 ~\forall t$. Using the full state evolution $\partial_t \hat{\rho}=\Lambda[\hat{\rho}]$, this further implies that SLS is avoided only for semi-group dynamics obeying the locality condition $\Tr_A [ \Lambda(\hat{\rho}) ]= 0$ at all times and for all states.

Again without loss of generality, we consider GKSL dynamics with only a single jump operator in Eq.~\eqref{GKSL}, of the form $\sqrt{\Gamma}\hat{L}=\hat{L}_A\otimes\hat{L}_B$. If we find SLS to be possible in this case, it will also be allowed in the more general case with multiple jump operators. Writing out $\Tr_A [ \Lambda(\hat{\rho}) ]$ explicitly yields:
\begin{align}
&\Tr_A [ \Lambda(\hat{\rho}) ] = \sum_{i,i',j,j'} 
 \vphantom{\ket{B}}^{\vphantom \dagger}_A\!\bra{i'}\hat{L}^\dagger_A \hat{L}^{\phantom\dagger}_A \ket{i}_A 
 \psi^{\phantom *}_{i,j} \psi^{\phantom *}_{i',j'} \times \notag \\
&~~~~~~\left( \hat{L}^{\phantom\dagger}_B \ket{j}_B \vphantom{\ket{B}}^{\vphantom \dagger}_B \!\bra{j'} \hat{L}_B^{\dagger} -\frac{1}{2}\left\{\hat{L}^{\phantom\dagger}_B \hat{L}_B^{\dagger}, \ket{j}_B \vphantom{\ket{B}}^{\vphantom \dagger}_B \!\bra{j'} \right\} \right)
\end{align}
The term in brackets is generically non-zero, and we thus find that linear GKSL dynamics in general allows for time dependence of Bob's reduced density matrix. Bob can then use instantaneous projective measurements to determine whether or not Alice initiated the DQSR dynamics, and thus establish superluminal signalling. 

If the jump operator $\hat{L}$ appearing in the GKSL evolution is local to Alice, so that $\hat{L}_B=\hat{\mathbb{I}}_B$, SLS is avoided. This may be used as a sufficient condition complementing the linearity of master equations to guarantee the absence of signalling, but notice that it is not a strictly necessary condition. More importantly, the condition that the collapse operator is local is distinct from, and not implied by, the condition that the state evolution of Alice's DQSR dynamics in Eq.~\eqref{state} acts only locally. To be certain whether or not a particular linear master equation of the GKSL form allows signalling, one thus has to explicitly evaluate its jump operators in a setting like that of Eq.~\eqref{state}, and ensure they either act locally, or always act in such a way that $\Tr_A[\Lambda(\rho)]=0$. Alternatively, the witness of Eq.~\eqref{martingale} provides a necessary and sufficient no-signalling condition in all circumstances.

An explicit example of local state evolution yielding non-local but still linear evolution of the full density matrix, is given by a simple toy-model in which Alice's device and state evolve deterministically. Consider the two-level dynamics of Eq.~\eqref{state}, acting on the (logical) basis states $\ket{M_A^i}\ket{i}_A\ket{i}_B\ket{M_B^i}$ for $i \in \{0,1\}$, and evolution according to $\partial_t \ket{\psi} = \gamma(\hat{\sigma}_A^z-\langle \hat{\sigma}_A^z \rangle) \ket{\psi}$. Here, the Pauli matrix $\hat{\sigma}_A^z = (\ket{M_A^0}\bra{M_A^0}-\ket{M_A^1}\bra{M_A^1})\otimes \mathbb{I}_A\otimes\mathbb{I}_B\otimes\mathbb{I}_{M_B}$ acts locally, on Alice's device only. The nonlinear term proportional to $\langle \hat{\sigma}_A^z \rangle$ ensures that normalisation is conserved, and does not affect the dynamics in any other way~\cite{Aritro_PRA}. The resulting deterministic dynamics of the density matrix can be solved exactly, and is written as $\hat{\rho}(t)=\ket{\psi(t)}\bra{\psi(t)}=[\,\hat{\tau}^0+\hat{\tau}^1/\cosh(2\gamma t)+\hat{\tau}^3 \tanh(2\gamma t)]/2$. Here $\hat{\tau}$ are Pauli operators acting on the full basis states of Alice and Bob combined, so that for example $\hat{\tau}^3=\ket{M_A^0} \ket{0}_A \ket{0}_B \ket{M_B^0}\bra{M_A^0} \bra{0}_A \bra{0}_B \bra{M_B^0}-\ket{M_A^1} \ket{1}_A \ket{1}_B \ket{M_B^1}\bra{M_A^1} \bra{1}_A \bra{1}_B \bra{M_B^1}$. The time evolution of the density matrix $\hat{\rho}(t)$ can be written as a linear master equation of the form of Eq.~\eqref{GKSL}, with time dependent coefficients. Explicitly, taking $\hat{H}=0$, the jump operators can be taken to be $\hat{L}_1=\hat{\tau}^3$ and $\hat{L}_2=\hat{\tau}^1$, with weights $\Gamma_1=\gamma\tanh(2\gamma t)$ and $\Gamma_2=\gamma(\tanh^2(2\gamma t)-1)/\tanh(2\gamma t)$. This map for the time evolution of $\hat{\rho}(t)$ can also be written in the form of a Kraus decomposition for small time increments: $\hat{\rho}(\delta t)=\sum_i\hat{K}_i\hat{\rho}(0)\hat{K}_i^{\dagger}$, with $\hat{K}_1=(\,\hat{\mathbb{I}}-\frac{\delta t}{2} (\Gamma_1 +\Gamma_2)\,\hat{\tau}^0$\,), $\hat{K}_2=\sqrt{\delta t\,\Gamma_1}\,\hat{\tau}^3$ and $\hat{K}_3=\sqrt{\delta t\,\Gamma_2}\,\hat{\tau}^1$. Written this way, the map is seen to be completely positive and trace preserving (CPTP), because $\sum_i\hat{K}_i^{\dagger}\hat{K}_i=\hat{\mathbb{I}}$, with $\hat{\mathbb{I}}$ the identity operator on the entire Hilbert space. Notice that the jump operator $\hat{L}_2$ can not be represented in terms of local operators acting only on Alice's device or state. We thus find that the evolution in this toy model allows for SLS, despite its state dynamics being generated by local operators and its ensemble dynamics obeying a linear, CPTP master equation. The presence of signalling is straightforwardly verified, as the dynamics reduces any initial state to $\ket{M_A^0}\ket{0}_A\ket{0}_B\ket{M_B^0}$ with unit probability.

\begin{figure}[tb]
\includegraphics[width=\columnwidth]{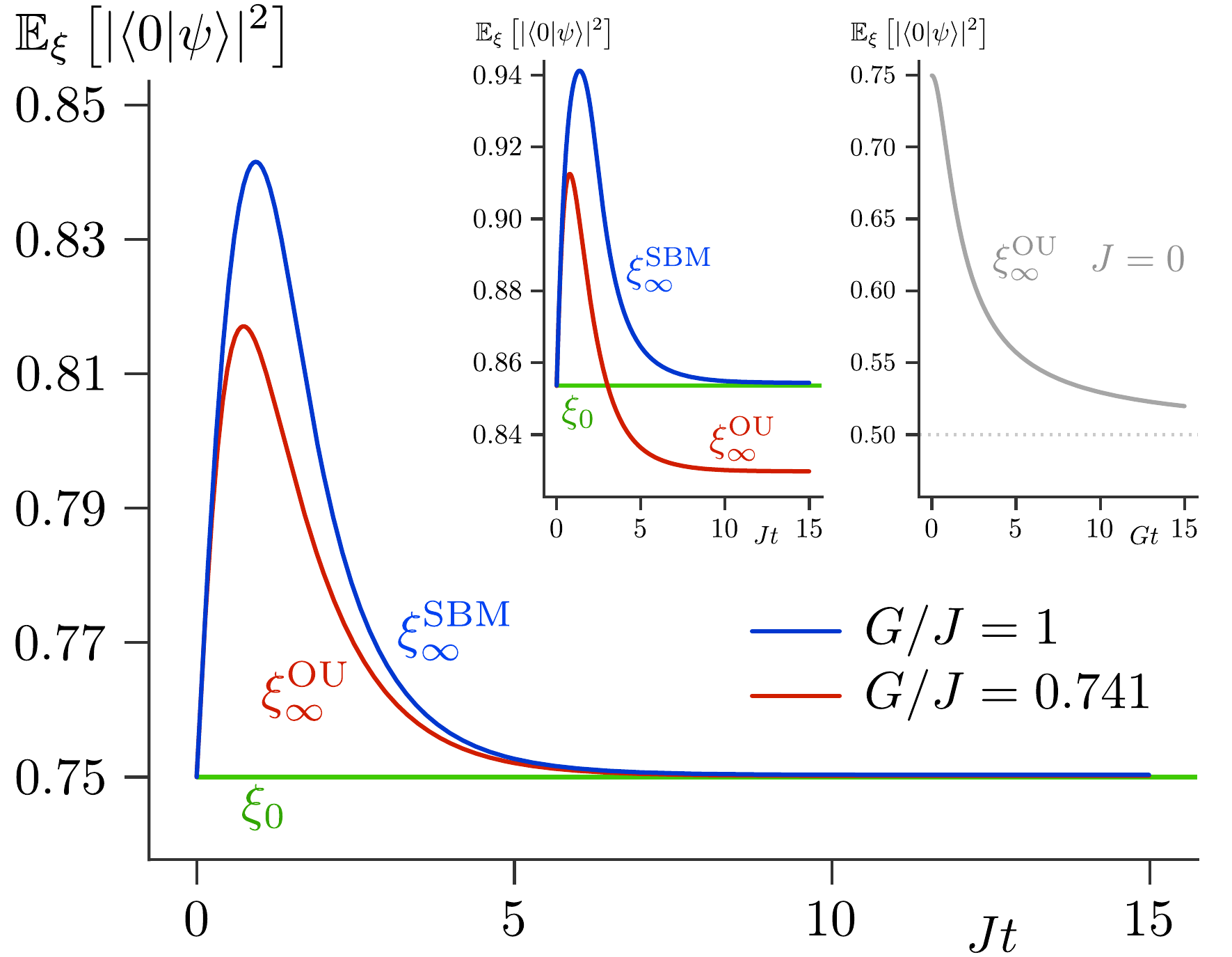}
\caption{\label{witnessfig} The quantity $\mathbb{E}_\xi\left[ \left| \alpha_{j=0}(\xi_t,t) \right|^2 \right]$ as a function of time for various DQSR models, evaluated using the dynamics specified in Eq.~\eqref{SUVdynamics}. Any deviations from the initial value signify a non-zero value for the witness of Eq.~\eqref{martingale}, and thus a possibility for superluminal signalling. The evolutions in the main panel all start from $\alpha(0)=\cos(\pi/6)$. A constant value of the witness indicates that the possibility of SLS is precluded. The green lines labeled $\xi_0$ show the limit of zero correlation time encountered in CSL as well as SUV models in the white-noise limit. The red and blue lines labeled $\xi_\infty$ indicate the opposite limit of infinitely correlated (constant) noise in the two-state SUV model, with noise values sampled either from the steady state of an Ornstein-Uhlenbeck (OU) process, or from spherical Brownian Motion (SBM). The left inset shows evolution with the same value for $G/J$, but for an initial state with $\alpha(0)=\cos(\pi/8)$. Comparing the main panel and the left inset shows that OU noise necessarily causes SLS even for long times, because for fixed ratio $G/J$ the witness does not return to its initial value for all possible initial conditions. In contrast, SBM noise does allow SLS to be avoided at long times. However, a non-constant value of the witness, and hence SLS, cannot be avoided at short times and constant noise values. The grey line in the right inset shows that a model with purely stochastic state evolution ($J=0$) will violate the no-signalling condition at all times. All expectation values shown are numerically averaged over $10^6$ implementations of the noise.}
\end{figure}

%%%%%%%%%%%%%%%%%

~\noindent\emph{\textbf{Applications of the witness condition} ---}
To demonstrate the application of the witness condition, we will explicitly evaluate it in a few particular instances. All DQSR models introduce weak perturbations to Schr\"odinger's equation and can be written in the form of coupled It\^o stochastic differential equations:
\begin{align}
    d\ket{\psi_t} &=\left( -i\,\hat{H}_t\,dt  + d\hat{\mathcal{G}}_t\right)\ket{\psi_t} \notag \\
    d\hat{\mathcal{G}}_t\ket{\psi_t} &= \hat{J}_t(\psi_t,\xi^\mu_t) \ket{\psi_t}dt + \sum_\nu \hat{G}^\nu_t(\psi_t,\xi^\mu_t) \ket{\psi_t}dW^\nu_t \notag \\
    d\xi^\mu_t &= f(\xi^\mu_t)dt + g(\xi^\mu_t)dW^\mu_t
    \label{DQSRgeneral}
\end{align}
Here $\hat{H}_t$ is the standard (possibly time dependent) Hamiltonian and $d\hat{\mathcal{G}}_t$ is the modification to time evolution defining the DQSR model. Its terms $\hat{J}_t$ and $\hat{G}^\nu_t$ may be (non-linear) functions of the state $\ket{\psi_t}$ and may further depend on a set of independent (possibly temporally correlated or colored) Markovian stochastic processes $\xi^\mu_t$ with continuous trajectories, indexed by $\mu\in \mathbb{Z}^+$. They follow independent It\^o stochastic differential equations (i.e. the noise average $\mathbb{E}_\xi[\xi_t^\mu \xi_s^\nu] = 0 ~\forall \mu\neq \nu$), in which $f$ and $g$ are smooth functions on their respective sample spaces~\cite{Aritro_Entropy,Aritro_PRA}. Here $dW^{\mu,\nu}_t$ are all independent identically distributed Gaussian increments denoting a standard Wiener process (corresponding to the Brownian motion $W_t=\int\,dW_t$)~\cite{revuz1999continuous,oksendal2003stochastic,gardiner2004handbook}.

The first class of DQSR models in this general formulation for which we consider the SLS witness is the well-known family of Continuous Spontaneous Localization (CSL) models~\cite{Pearle_89_PRA, Bassi_03_PhyRep,Ghirardi_90_PRA, percival1998}. Applied to an initial two-state superposition, they are defined by:
\begin{align*}
d\hat{\mathcal{G}}_t \ket{\psi_t} &= \left[\frac{-\gamma}{2}\left(\hat{\sigma}-\langle\hat{\sigma}\rangle_t\right)^2 dt +  \sqrt{\gamma}\left(\hat{\sigma}-\langle\hat{\sigma}\rangle_t\right)dW_t\right] \ket{\psi_t}
\end{align*}
Here $\gamma$ is a constant and $\hat{\sigma}$ is a Pauli-$\sigma_z$ operator. We consider the setup described before where the initial state contains products of the states of local measurement machines of Alice and Bob as well as a shared entangled state. The Pauli operator is then defined in terms of the states $|M_A^{0,1}(t)\rangle$ describing Alice's measurement device. Notice that this model has only uncorrelated (white) noise described by $dW_t$. The recently introduced family of models for DQSR known as Spontaneous Unitarity Violation (SUV)~\cite{symmetry,vanwezelprb,mertenspra,Aritro_Entropy,Aritro_PRA,Aritro_Arxiv}, also have a white-noise limit that falls in this class of CSL models~\cite{Aritro_PRA,Aritro_Arxiv}.

For these white-noise models, a GKSL-type master equation can be formulated, obeying the Martingale property for its probability dynamics~\cite{Bassi_03_PhyRep,Aritro_PRA}. The witness of Eq.~\eqref{martingale} is then analytically verified to be zero for all time, precluding the possibility of SLS. This is indicated by the green line in Fig.~\ref{witnessfig}, where $\xi_0$ indicates the vanishing correlation time of noise in CSL models. 

Going beyond white-noise models, non-Markovian quantum state diffusion has been considered in the contexts of open quantum system dynamics~\cite{Diosi98_non_markov,Diosi99_non_Markov_open}, and of DQSR models driven by temporally correlated Gaussian noise ~\cite{cCSL1,cCSL2}. In these cases it is typically not possible to find a closed form expression for the dynamics without restricting to particular settings~\cite{Bassi_03_PhyRep,cCSL1,Diosi98_non_markov}. In the cases that a linear master equation can be obtained for the ensemble dynamics, this is again not sufficient to guarantee the absence of SLS. For instance, the master equation found in Refs.~\cite{Bassi_03_PhyRep,cCSL1} for a particular family of non-Markovian models is of the linear form $\partial_t \hat{\rho} = \tilde{\Lambda}_t[ \hat{\rho}] = -\gamma\sum_{\mu,\nu} \int_0^t ds\, \mathcal{D}_{\mu\nu}(t,s) [\hat{L}_\mu,[\hat{L}_\nu,\hat{\rho}(t)]]$. The GKSL form of Eq.~\eqref{GKSL} can be seen to be a special case of this dynamics, and therefore all arguments showing that GKSL type equations allow SLS, also apply to this more general form. As before, an additional assumption on the jump operators and memory kernel is required to guarantee that $Tr_A [ \tilde{\Lambda}_t(\hat{\rho}) ]=0$ in setups with spatially separated actors. Similar reasoning also applies to the types of non-Markovian dynamics proposed on the basis of different assumption in Refs.~\cite{Bassi_03_PhyRep,cCSL1,cCSL2,Diosi98_non_markov}, as well as in relativistic DQSR models~\cite{Bassi_03_PhyRep,Bassi_rela_Jones_2021}. Each time, the vanishing of the witness condition of Eq.~\eqref{martingale} does not follow from linearity of the master equation or the locality the state evolution, but needs to be ascertained for individual instances of Markovian and non-Markovian DQSR models in order to ensure the absence of signalling.

Going beyond situations where a linear master equation may be obtained analytically, one of the particular advantages of the witness condition in Eq.~\eqref{martingale} is that it can also be numerically evaluated in situations where analytical expressions of the master equations are not known. This applies in particular to models with correlated noise that have been considered in the context of SUV, for which no closed-form master equation is available~\cite{mertenspra,mertensscipost,mertenscore,Aritro_Entropy}. We will numerically evaluate the witness condition of Eq.~\eqref{martingale} for two of these models, and show that they too require further restrictions to avoid SLS.

\begin{figure}[tb]
\includegraphics[width=\columnwidth]{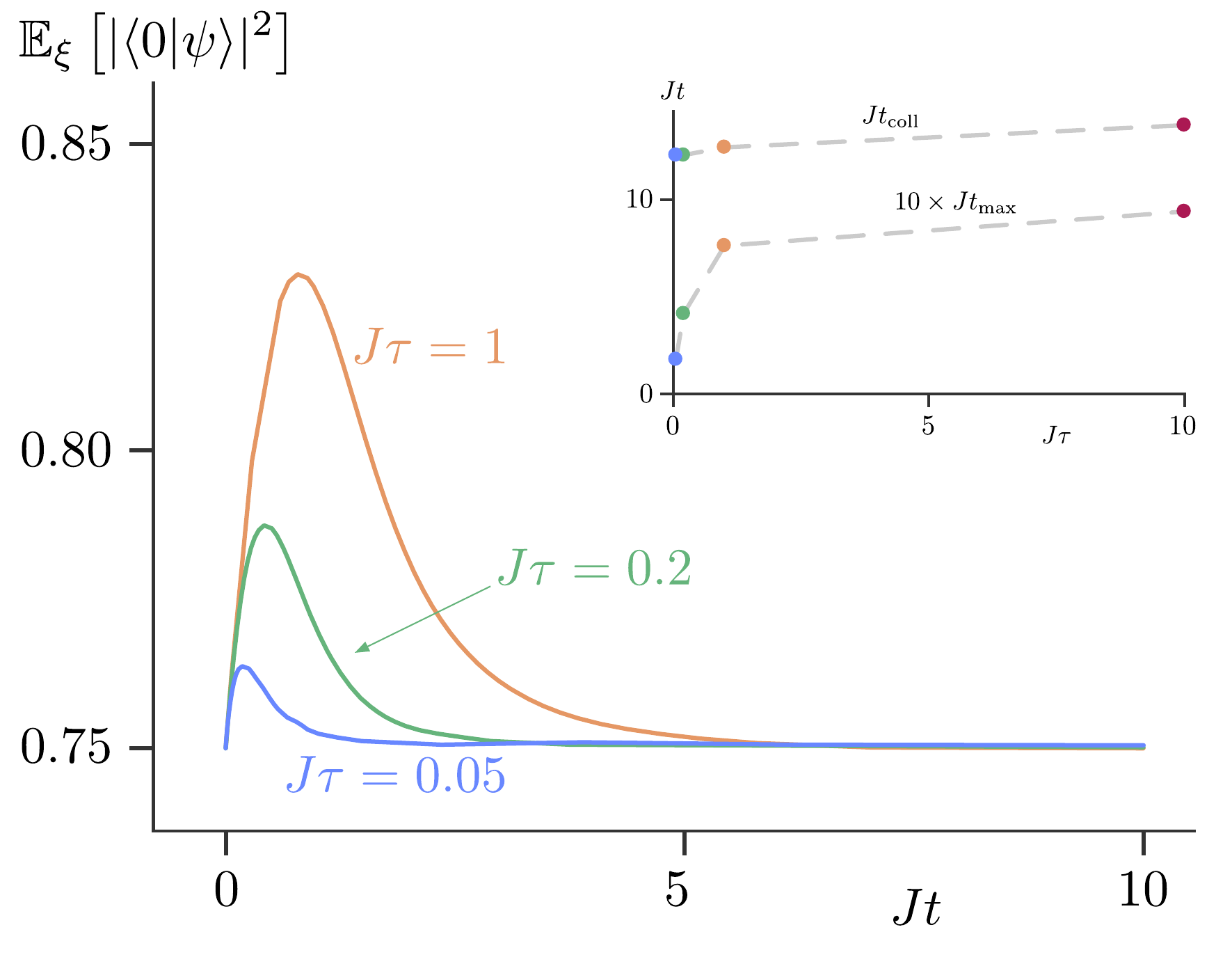}
\caption{\label{bumpfig} The quantity 
$\mathbb{E}_\xi\left[ \left| \alpha_{j=0}(\xi_t,t) \right|^2 \right]$ as a function of time for the SUV model with correlated SBM noise. Any deviations from the initial value signify a non-zero value for the witness of Eq.~\eqref{martingale}, and thus a possibility for superluminal signalling. The lowest (blue) curve is for correlation time $J\tau=0.05$, the middle (green) one for $J\tau=0.2$, and the top (orange) line for $J\tau=1$. For each evolution, the value of $J/G$ was adjusted to ensure the emergence of Born's rule at long times. The inset shows that the time $Jt_{\text{max}}$ at which the maximum deviation of the witness from its initial value occurs, shifts to later times for larger correlation times. Despite this, it always stays well below the collapse time $Jt_{\text{coll}}$, defined as the characteristic decay time of the off-diagonal expectation value $\mathbb{E}_\xi\left[ \alpha_{j=0}(\xi_t,t) \alpha_{j=1}(\xi_t,t) \right]$. All expectation values shown are numerically averaged over $10^5$ implementations of the noise.}
\end{figure}

For two-state superpositions, SUV dynamics is described by:
\begin{align}
d\hat{\mathcal{G}}_t\ket{\psi_t} &= \left(J\langle\hat{\sigma}\rangle_t + G \xi_t \right) \left(\hat{\sigma}-\langle\hat{\sigma}\rangle_t\right) \ket{\psi_t}dt 
\label{SUVdynamics}
\end{align}
Here $\hat{\sigma}$ again indicates a Pauli-$\sigma^z$ operator acting on Alice's measurement device. The overall geometric factor introduced by the final expectation value ensures norm-preserving dynamics, while the non-linear term proportional to $J$ induces stable end points for the evolution at $\langle\hat{\sigma}\rangle=0$ or $1$, corresponding to the two possible measurement outcomes~\cite{Aritro_PRA,mertenspra}. The stochastic parameter $\xi_t$ is assumed to be temporally correlated, and obeys the independent stochastic dynamics $d\xi_t=-\xi_t dt/\tau + g(\xi_t)dW_t$. Here, $\tau$ is the correlation time for the noise. Although any generic continuous stochastic process may be utilized for $\xi_t$, two processes of particular interest are the (Gaussian) Ornstein-Uhlenbeck process (OU) with $g(\xi_t)=\sqrt{{2}/{\tau}}$ and the spherical Brownian motion (SBM) defined by $g(\xi_t)=\sqrt{{1-\xi_t^2}/{\tau}}$~\cite{mertenspra,mertensscipost,mertenscore,Aritro_PRA}. 

In the limit of infinite correlation time (static noise), the results of numerically evaluating Eq.~\eqref{SUVdynamics} is depicted by the red and blue lines in Fig.~\ref{witnessfig}. For SBM noise, it is possible to find a ratio $G/J$ such that Born's rule is obeyed at long times for all possible initial conditions. This precludes SLS in the long time limit. At short times, however, there is a clear deviation from $\mathbb{E}_\xi[ | \alpha_{j=0}(\xi_t,t) |^2]$ being constant, and the witness condition of Eq.~\eqref{martingale} thus implies that the possibility of SLS is unavoidable. Moreover, for OU noise, SLS is unavoidable even at long times, because there is no value of $G/J$ that allows the witness to return to its initial value for all initial states simultaneously~\cite{mertenscore}.

For intermediate values of the correlation time in SUV models with SBM noise, Fig.~\ref{bumpfig} shows that the short-time deviation of $\mathbb{E}_\xi\left[ \left| \alpha_{j=0}(\xi_t,t) \right|^2 \right]$ from its initial value disappears smoothly in the limit of vanishing correlation time, but cannot be avoided for any finite value. This has a direct implications in the limit of large measurement devices, as the time scale of DQSR dynamics decreases with system size. In SUV models, for example, this is due to the extensive scaling of the parameter $J$ with system size. Moreover, all physical noise process have non-zero correlation time and regardless how small it is, there will be a critical size for measurement machines above which the DQSR evolution takes place entirely within the correlation time of the noise. SLS is then unavoidable for macroscopic measurement machines obeying the types of DQSR dynamics considered above. Notice however, that although the time at which the maximum deviation occurs shifts to higher values for longer correlation times of the noise, it stays well below the characteristic time over which the evolution approaches its stable end points (as shown in the inset of Fig.~\ref{bumpfig}). The unavoidable possibility of SLS occurring in macroscopic devices, indicated by the witness deviating from a constant value, therefore occurs within a vanishingly short time-interval in the thermodynamic limit, and may not be accessible or observable in practice.

That SLS cannot be avoided in the models considered in principle, even if its effects are unobservable in practice, should raise questions about the existence of closed time-like loops. Notice however, that while we did not identify any DQSR processes with correlated noise and without analytically obtainable master equations that are capable of avoiding SLS in general, it is possible that particular signalling-free instances of such processes exist. If so, they can be straightforwardly identified by the witness introduced here. A possible place to look for such processes is the generalisation of SUV processes to initial configurations superposed over a continuous rather than a discrete set of states.

%%%%%%%%%%%%%%%%%%%%

~\noindent\emph{\textbf{Conclusions} ---}
In conclusion, we showed that the witness conditions of Eqs.~\eqref{martingale} and~\eqref{generalMartingale} provide both a necessary and sufficient condition for any model introducing changes to Schr\"odinger's equation to preclude the possibility of superluminal signalling. The condition ensures that the marginalized probability dynamics is a Martingale process. It applies to modifications of quantum theory in general, and in particular to models for dynamical quantum state reduction.

Applying the witness condition to models whose ensemble dynamics is described by a GKSL-type master equation, we find that linearity of the master equation is a necessary but not a sufficient condition to rule out superluminal signalling. This applies both to Markovian and non-Markovian models. We showed that in these cases, one possible way to guarantee the absence of signalling, is to supplement the linearity condition on the master equation by a locality condition on its jump operators. We stress, however, that locality of the jump operators is not implied by locality of the state evolution, and has to be confirmed in each individual microscopic model. Alternatively, the witness condition of Eq.~\eqref{martingale} provides a necessary and sufficient condition for certifying the absence of superluminal signalling under all circumstances. 

Going beyond descriptions in terms of ensemble dynamics, the witness condition of Eq.~\eqref{martingale} can also be straightforwardly checked for models in which analytical expressions of the master equations are not known. We gave explicit examples of this for specific models with correlated noise, which we showed to allow superluminal signalling at short times, even if they conform to Born's rule at long times. This does not imply signalling-free models driven by colored noise (with or without linear master equations) do not exist. In fact, we argue that if dynamical state reduction, or any other type of modified Schr\"odinger dynamics, is caused by a physical process, it cannot be fundamentally white and will have to obey Eq.~\eqref{martingale} in order to avoid superluminal signalling. The condition formulated here thus provides a stringent requirement as well as a clear guiding principle in the ongoing search for a fully satisfactory model of dynamical quantum state reduction.

%%%%%%%%%%%%%%%%%%%%%%%%%%%%%%%%

\subsection*{Acknowledgement}
The authors gratefully acknowledge illuminating discussions with  A. Bassi, L. Diosi, D. Snoke and H. Maassen. A.M also thanks S. M. P. Devi and J\~n$\bar{\mathrm{a}}$nananda Seva Sangha for their gracious support.

%%%%%%%%%%%%%%%%%%%%%%%%%%%%%%%%

%

%%%%%%%%%%%%%%%%%%%%%%%%%%%%%%%%

\end{document}